\documentclass[graybox,natbib,nosecnum]{svmult}
\bibpunct{(}{)}{;}{a}{}{,} 

\pdfoutput=1   

\usepackage{mathptmx}       
\usepackage{helvet}         
\usepackage{courier}        
\usepackage{type1cm}        

\usepackage{makeidx}         
\usepackage{graphicx}        
\usepackage{multicol}        
\usepackage[bottom]{footmisc}
\usepackage[normalem]{ulem}	
\usepackage{hyperref}  

\usepackage{soul}   

\newcommand{\msun}{\, {\rm M}_{\odot}}
\newcommand{\peryear}{\, {\rm yr}^{-1}}

\makeindex             

\begin{document}

\title*{Planetary Migration in Protoplanetary Disks}
\author{Richard P. Nelson}
\institute{Richard P. Nelson \at Astronomy Unit, Queen Mary University of London, Mile End Road, London, E1 4NS, U.K., \email{R.P.Nelson@qmul.ac.uk}
}
%
%
\maketitle

\abstract{The known exoplanet population displays a great diversity of orbital architectures, and explaining the origin of this is a major challenge for planet formation theories. The gravitational interaction between young planets and their protoplanetary disks provides one way in which planetary orbits can be shaped during the formation epoch. Disk-planet interactions are strongly influenced by the structure and physical processes that drive the evolution of the protoplanetary disk. In this review we focus on how disk-planet interactions drive the migration of planets when different assumptions are made about the physics of angular momentum transport, and how it drives accretion flows in protoplanetary disk models. In particular, we consider migration in discs where: (i) accretion flows arise because turbulence diffusively transports angular momentum; (ii) laminar accretion flows are confined to thin, ionised layers near disk surfaces and are driven by the launching of magneto-centrifugal winds, with the midplane being completely inert; (iii) laminar accretion flows pervade the full column density of the disc, and are driven by a combination of large scale horizontal and vertical magnetic fields.}

\section{Introduction }
The realisation that disk-planet interactions lead to the exchange of angular momentum between the planet and disk, leading to the possibility of planetary migration and the shaping of the disk, predates the discovery of extrasolar planets by more than a decade \citep[e.g.][]{GoldreichTremaine1979,GoldreichTremaine1980,LinPapaloizou1979}. Since this time, the discovery of extrasolar planets with a great variety of orbital architectures, ranging from hot Jupiters with periods of a few days \citep{MayorQueloz1995}, compact systems of terrestrial planets and super-Earths \citep[e.g.][]{Lissauer2011}, and long-period super-Jovian planets orbiting at distances of tens of AU \citep[e.g.][]{Marois2008} has stimulated an enormous amount of theoretical work aimed at understanding the basic mechanisms involved in disk-driven planet migration, and how migration shapes the final orbits of planetary systems as it operates during the epoch of formation.

The discovery of short period exoplanets was initially believed to provide evidence for large scale planet migration \citep[e.g.][]{Lin1996}. In recent years, however, in situ formation models have been proposed as a means of explaining observed planetary systems \citep{HansenMurray2012}. One of the key early predictions of migration theory was that multiplanet systems ought to show a preference for being in mean motion resonances \citep[e.g.][]{CresswellNelson2006}, and while a number of resonant systems have been discovered such as GJ 876 \citep{Marcy2001}, Kepler 223 \citep{Mills2016} and the Trappist-1 system \citep{Gillon2017}, the number of bona fide resonant systems present among the 100s of multiplanet systems discovered by NASA's Kepler mission is below original expectations. It is unclear at present if this is best explained by some process operating that breaks resonances \citep[e.g.][]{Goldreich2014,Izidoro2017}, or if instead resonance formation is a relatively rare process. Nonetheless, the existence of some resonant systems would seem to show beyond reasonable doubt that migration is important during planet formation, and in situ models then need to explain why net migration torques due to disk-planet interactions are expected to be ineffective at driving planet migration. Further compelling evidence for migration is provided by the circumbinary planets orbiting very close to the zones of instability around close binary systems, such as Kepler 16, Kepler 34 and Kepler 35 \citep[e.g.][]{Doyle2011}, since this environment would be very challenging for in situ planet formation. The question would seem to be not whether or not migration occurs during planet formation, but rather how important is it in shaping the final planetary systems versus other effects such as the radial drift of planet-building solids through aerodynamic drag, etc?

A number of comprehensive reviews of disk-planet interactions have been published in recent years \citep{KleyNelson2012,Baruteau2014}. In this shorter review, we have focussed on the question of how a single planet on a circular orbit migrates, and influences the local disk structure, when embedded in different disk models that are characterised by the physical processes that drive radial accretion flows in the disk. This approach is motivated by the fact that our understanding of protoplanetary disk dynamics has changed considerably over recent years, because of developments in both observations and theory. Whereas early theoretical work on protoplanetary disk evolution assumed that angular momentum transport in disks arose because of an effective viscosity provided by turbulence, recent non-ideal MHD simulations point to accretion occurring in the surface layers of completely laminar disks, or possibly in their midplanes, driven by large scale magnetic fields. As discussed below, the nature of the disk model strongly influences the outcome of disk-planet interactions. 

Observations also provide important clues to the main processes driving angular momentum transport in discs. Recent attempts to detect turbulent motions in the outer regions of the protoplanetary discs HD 163296 and TW Hydra with ALMA, where non-ideal MHD simulations suggest that turbulent motions might be present \citep{Simon2015}, have not made definitive detections and have instead only been able to place upper limits on the turbulent velocities \citep{Flaherty2015,Teague2016}, suggesting that any turbulence there is relatively weak. Highly collimated jets, and large scale molecular outflows, are observed almost ubiquitously around young stars, and these are thought to be launched and collimated by large scale magnetic fields threading the surrounding protoplanetary disk \citep[see the review by][and references therein]{Frank2014}. Specific examples include the famous jet emerging from the inner regions of the T Tauri star HH30  
\citep{Burrows1996} and its accompanying molecular outflow \citep{Pety2006}, and the rotating outflow associated with HH212 discovered recently with ALMA \citep{Tabone2017}.  These jets and outflows are generally believed to be magneto-centrifugally launched from the protoplanetary disc \citep{BlandfordPayne1982, PudritzNorman1983}, leading to loss of angular momentum from the disc, and the observational detection of rotation in the outflows of \emph{some} these systems provides supporting evidence for this model. While more observations are clearly required before we have a good understanding of the angular momentum transport mechanisms that drive accretion in protoplanetary discs, there may be an emerging convergence between theoretical expectations and observational evidence for the importance of large scale magnetic fields and magneto-centrifugal wind launching. 

\section{Protoplanetary disk models}
The orbital evolution of a planet embedded in a protoplanetary disk depends qualitatively and quantitatively on the structure of the disk, and the physical processes that govern its evolution. When we say \emph{embedded in a protoplanetary disk}, what we actually mean is \emph{embedded in a protoplanetary disk model}, because, in spite of the spectacular progress that has been made in observationally characterising protoplanetary disks in recent years, including the discovery of rings and other features in ALMA images of systems such as HL Tau \citep{ALMA2015} and TW Hydra \citep{Andrews2016}, which have been interpreted as being due to an array of different phenomena (see chapter in this volume by Andrews \& Birnstiel), observational data are still unable to provide a complete picture of disk structure, or of the processes that drive disk evolution. Hence, analyses of planet migration are based on theoretical disk models, and these can include a wide array of different physical processes that substantially affect predictions of how planets migrate. For this reason, our discussion about disk-planet interactions is organised according to the type of disk model under consideration. In this article, we consider three distinct disk models, that we refer to as: \emph{viscous disks}; \emph{inviscid disks}; \emph{advective disks}. This terminology is explained below.

Young stars are observed to accrete gas from their surrounding protoplanetary disks  at a canonical rate of $\sim 10^{-8} \msun \peryear$ \citep[e.g.][]{Hartmann1998}, and are observed to have finite lifetimes of a few Myr \citep{Haisch2001,Hillenbrand2008}, indicating that some process leads to the evolution of the disk through the removal of angular momentum so that gas can flow radially through the disk and onto the central star. 

Work undertaken during the early development of accretion disk theory suggested that turbulence in the disk, of unknown origin, gives rise to an anomalously large effective viscosity that causes diffusive transport of angular momentum \citep{ShakuraSunyaev1973,LyndenBellPringle1974}. The internal viscous transport of angular momentum in a disk causes most of the mass to flow inwards onto the central star, with a small amount of gas absorbing the angular momentum and drifting out to large radius. In this work, we define a disk model in which angular momentum transport occurs primarily through viscous/turbulent diffusion throughout the body of the disk to be a \emph{viscous disk}.

Turbulence can be generated in a disk containing a weak magnetic field through the magneto-rotational instability (MRI) \citep{BalbusHawley1991}, but calculations suggest that the MRI is not capable of transporting angular momentum in the main body of a protoplanetary disk, particularly at the midplane, because the ionisation fraction of the gas is too small to provide good coupling between it and the magnetic field \citep{BlaesBalbus1994}. Regions that may be capable of coupling strongly to magnetic fields include the inner few tenths of an AU, where the temperature may be high enough for alkali metals such as potassium and sodium to be thermally ionised \citep{DeschTurner2015}, and in thin layers at the disk surface where external sources of ionisation such as stellar X-rays, UV photons and cosmic rays are able to penetrate \citep{Gammie1996,Perez-BeckerChiang2011,Glassgold1997}. This gives rise to a picture of protoplanetary disks in which accretion may occur only in the surface layers over a large range of stellocentric distances, with the midplane maintaining a \emph{dead zone} in which the flow remains laminar \citep{Gammie1996}. 

Recent work shows that when non-ideal MHD processes (i.e. Ohmic resistivity, ambipolar diffusion, the Hall effect) are included in MHD simulations of disks that include ionisation chemistry and a relatively weak vertical magnetic field, then there are two main outcomes. When the magnetic field is anti-aligned with the disk angular velocity vector, then the Hall effect plays only a minor role, and Ohmic resistivity and ambipolar diffusion quench any MRI turbulence, but allow a centrifugally driven disk wind to be launched from the ionised surface layers \citep{BaiStone2013,Gressel2015}. The launching mechanism for magneto-centrifugal winds had been examined much earlier, in application to protoplanetary disc models where gas could be launched into the wind from much closer to the midplane, so it has been understood for a long time that such a model could, in principle, explain the observed jets and outflows associated with young stars, as well as accretion onto the central star \citep{BlandfordPayne1982, PudritzNorman1983,PudritzNorman1986}. The launching mechanism in non-ideal discs involves transfer of energy and angular momentum from thin surface layers to the gas being launched into the wind, and hence the surface layers accrete towards the star. The simulations of \cite{BaiStone2013} and \cite{Gressel2015} give accretion rates in good agreement with the observations. In this work, we refer to a disk model in which the disk remains entirely laminar, but accretion occurs in thin surface layers due to wind launching, as an \emph{inviscid disk}, since there is no angular momentum transport or gas flow near the midplane where planets may be located. A key (and untested) assumption here is that the flow in the surface layers does not affect disk-planet interactions that may be occurring near the midplane.

When the magnetic field is aligned with the angular velocity vector of the disk, then the Hall effect becomes important. Now the Hall-shear instability \citep{Kunz2008} can generate radial magnetic fields that diffuse towards the disk midplane, and wind up due to the Keplerian shear to form strong horizontal fields that generate a significant accretion stress there \citep{KunzLesur2013,LesurKunzFromang2014,Bai2014,Bethune2017, Bai2017}. While we should note that the most recent, state-of-the-art defining simulations by \cite{Bethune2017} and \cite{Bai2017} require some improvement before they can be said to provide a realistic picture of protoplanetary disk evolution, they do indicate there are circumstances where the disk remains entirely laminar, and accretion in the surface layers due to wind launching is accompanied by accretion in the midplane due to the winding up of Hall-generated magnetic fields. In this work, we refer to a disk model in which laminar accretion flows occur throughout the body of the disk due to magnetic stresses as an \emph{advective disk}, in contrast to the diffusive transport of angular momentum that occurs in a viscous disk.

Throughout this article we will make numerical estimates of various quantities, such as the expected migration rates of planets. These require specification of a disk model, as they depend on quantities such as the local surface density, $\Sigma$ and temperature, $T$, and for simplicity we adopt radial power-law distributions for these quantities: \begin{equation}
\Sigma(r) = \Sigma_0 \left( \frac{r}{1 \, {\rm AU} } \right)^{-\alpha}, \;\;\; T(r) = T_0 \left(\frac{r}{1 \, {\rm AU}}\right)^{-\beta}.
\label{eqn:Sigma+T}
\end{equation}
In general we expect $\alpha$ and $\beta$ to be positive numbers of order unity.  We adopt parameters from the minimum mass solar nebula (MMSN) model of \cite{Hayashi1981} at an orbital radius of 1 AU when defining the values of $\Sigma_0$ and $T_0$:
\begin{equation}
\Sigma_0  =  1700 \; {\rm g} \; {\rm cm}^{-2}, \;\;\; T_0 =  270 \; {\rm K}, \;\;\; c_{\rm s,0}  =  1 \; {\rm km} \; {\rm s}^{-1},
\label{eqn:MMSN}
\end{equation}
where $c_{\rm s,0}$ is the isothermal sound speed. An estimate of the ratio of the disk pressure scale height (or ``thickness" $H$) to the orbital radius at 1 AU is then given by $h\equiv H/r=(c_{\rm s,0}/v_{\rm k}) \sim 0.035$, where $v_{\rm k}$ is the Keplerian velocity. Note that $h$ will refer to the aspect ratio of the disk, as defined above, throughout this article.

When considering viscous disk models, we note that for a steady disk the mass accretion rate and kinematic viscosity are given by \citep{Pringle1981}
\begin{equation}
{\dot M} \simeq 3 \pi \nu \Sigma,  \;\;\;\;\; \nu = \alpha_{\rm ss} c_s H
\label{eqn:mdot+nu}
\end{equation}
where the second expression is the so-called alpha prescription for the kinematic viscosity \citep{ShakuraSunyaev1973}, and $\alpha_{\rm ss}$ is a parameter that defines the strength of the viscous/turbulent diffusion of angular momentum. For a canonical accretion rate of ${\dot M}=10^{-8} \msun \peryear$, we obtain $\alpha_{\rm ss} \sim 10^{-3}$ for a typical \emph{viscous disk model} using the above disk parameters at 1 AU.

\section{Migration in viscous disks}
For viscous disk models, the response of the disk to the presence of the planet depends sensitively on the mass of the planet, because viscosity can be effective at smoothing perturbations induced by a planet. Hence, we discuss low mass, intermediate mass and high mass planets separately below.
\subsection{Low mass planets}
We define a low mass planet to be one that does not significantly alter the local disk structure. As discussed below, what constitutes a low planet according to this definition depends on what type of disk model one is considering. For a \emph{viscous disk model} with parameters similar to those described above, a low mass planet can have a mass up to $\sim$ few tens of Earth masses.

The gravitational interaction between a planet and the disk in which it is embedded leads to a torque acting on the planet (and on the disk because of angular momentum conservation) that is normally considered to have two components: the \emph{Lindblad torque} and the \emph{corotation torque}. The sum of these determines the direction and speed of the planet's migration. If a net torque, $\Gamma$, is applied to a planet, then it will migrate at a rate given by
\begin{equation}
\frac{d r_{\rm p}}{dt} = 2\frac{\Gamma}{m_{\rm p}} \sqrt{\frac{r_{\rm p}}{GM_*}},
\label{eqn:dadt}
\end{equation}
where $m_{\rm p}$ and $r_{\rm p}$ are the planet mass and orbital radius, respectively, and $M_*$ is the mass of the central star. Equation~(\ref{eqn:dadt}) can be obtained simply by differentiating with respect to time the expression $J_{\rm p} = m_{\rm p} \sqrt{GM_* r_{\rm p}}$, where $J_{\rm p}$ is the angular momentum of a planet on a circular Keplerian orbit, noting that ${\dot J}_{\rm p} \equiv \Gamma$. 
\subsubsection{Lindblad torque}
A planet embedded in a gaseous disk launches density waves at Lindblad resonances \citep{GoldreichTremaine1979,GoldreichTremaine1980}. These resonances correspond to locations in the disk where the forcing frequency due to the planet, as seen by a fluid element orbiting in the disk, equals the epicyclic frequency of the fluid element, $\kappa$. The epicyclic frequency is the natural radial oscillation frequency of a particle orbiting in the disk, and so we see that the system acts like a forced oscillator: fluid elements shear past the planet on circular orbits, and those elements that orbit at the radial locations of the Lindblad resonances undergo large amplitude radial oscillations that result in the excitation of acoustic waves that propagate away from the resonances. As these waves propagate radially, the Keplerian shear causes them to appear as spiral density waves. 

Working in cylindrical polar coordinates ($r$, $\phi$), the gravitational potential of a planet on a circular orbit can be expressed as a Fourier series
\begin{equation}
\Phi_{\rm p}(r, \phi, t) \equiv - \frac{G m_{\rm p}}{| {\bf r}_{\rm p} - {\bf r} |} = \sum_{m=0}^{\infty} \Phi_m(r) \cos{\left(m \left[ \phi - \phi_{\rm p} \right] \right)},
\label{eqn:phi_p}
\end{equation}
where $\phi_{\rm p}= \Omega_{\rm p} t$ is the azimuthal angle of the planet, moving on a circular orbit with angular velocity $\Omega_{\rm p}$,
and $\Phi_m(r)$ are the Fourier coefficients corresponding to each azimuthal mode number $m$. Hence, we see that the planet potential is represented as a superposition of terms, each with $m$-fold azimuthal symmetry, that rotate with a pattern speed equal to the planet's orbital angular velocity, $\Omega_{\rm p}$, when viewed in the inertial frame. When viewed in a frame moving with a fluid element orbiting in the disk at radius $r$, however, the frequency of each mode seen by the fluid element is now $m (\Omega_{\rm p} - \Omega(r))$. Equating the forcing frequency of each mode with the epicyclic frequency at locations in the disk that sit both interior and exterior to the planet (noting that $\kappa(r)=\Omega(r)$ in a Keplieran disk), gives the following expression 
\begin{equation}
\Omega(r) = \frac{m}{m \pm 1} \Omega_{\rm p},
\label{eqn:Omega_LR}
\end{equation}
corresponding to the locations of the outer (+) and inner (-) Lindblad resonances.
We see that the outermost $m=1$ Lindblad resonance occurs where $\Omega(r)=\Omega_{\rm p}/2$. The innermost $m=2$ Lindblad resonance occurs where $\Omega(r)=2 \Omega_{\rm p}$.
\begin{figure}
\includegraphics[scale=.25]{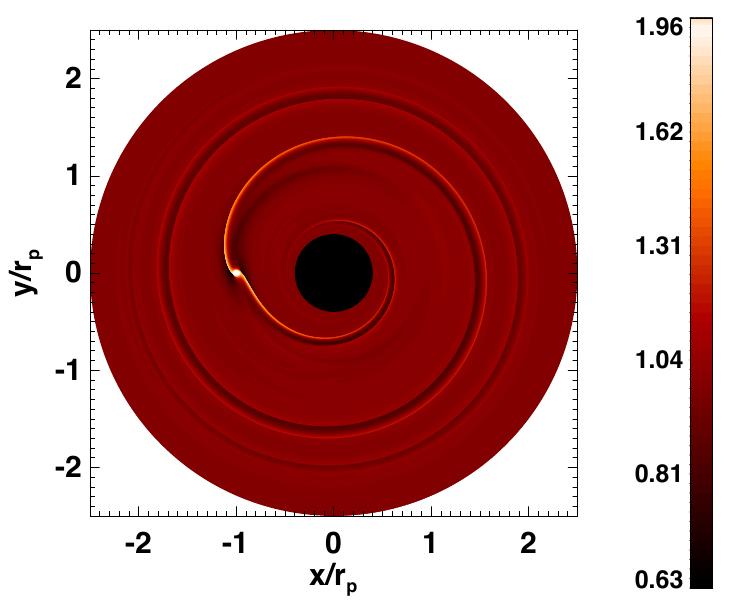}
\includegraphics[scale=.4]{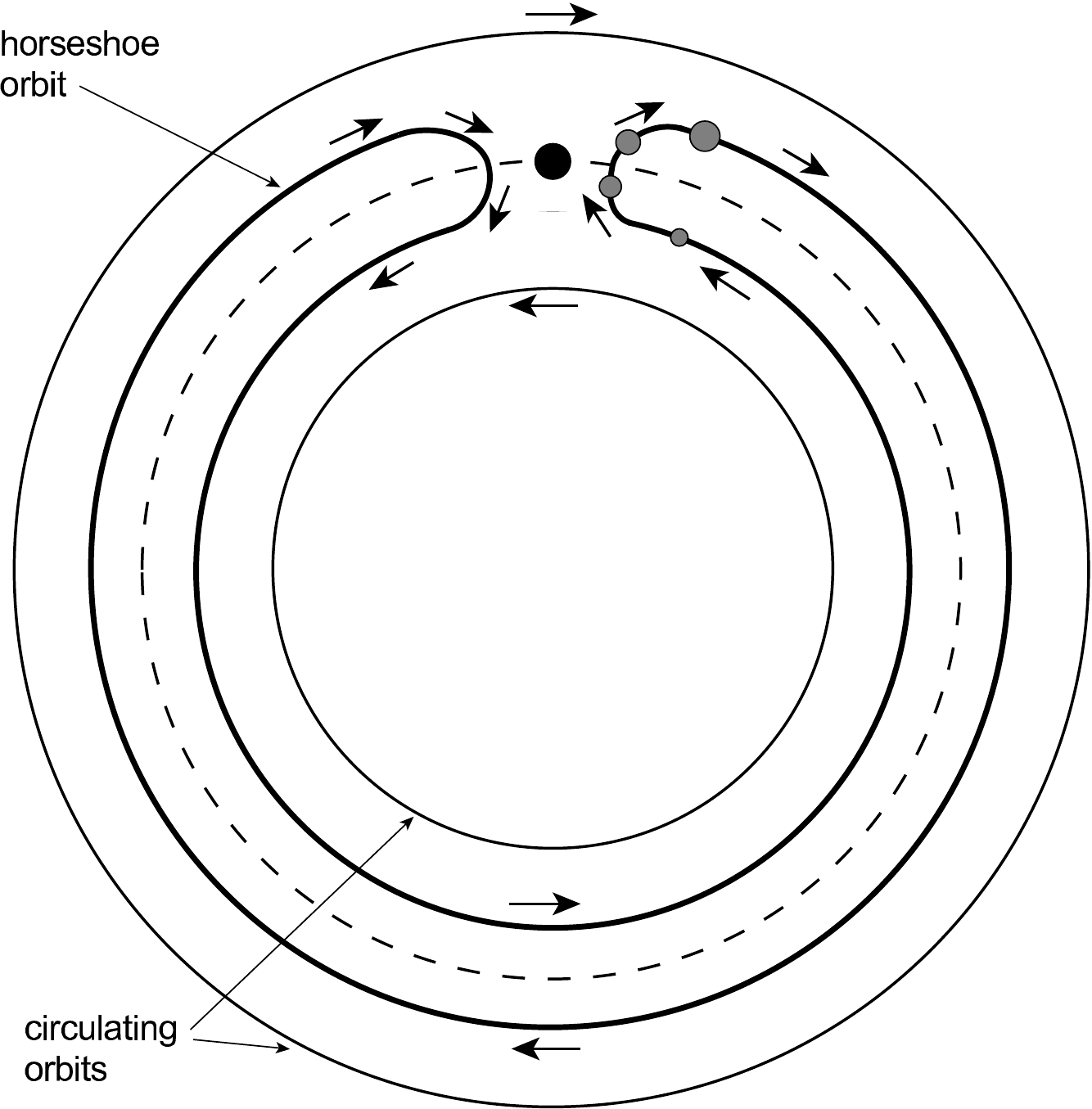}
\caption{Left panel: spiral waves generated by a 10 M$_{\oplus}$ planet embedded in a viscous disk. Right panel: schematic diagram showing horseshoe orbits with respect to a planet taken from \citet{Armitage2007}.}
\label{fig:spiral-horseshoe}       
\end{figure}

We now have a picture in which spiral waves with $m$-fold symmetry are excited at each of the Lindblad resonances, and propagate away from them at approximately the sound speed. Each wave has a pattern speed $\Omega_{\rm p}$, the orbital angular velocity of the planet. The waves carry an angular momentum flux into the disk, and their angular momentum content is deposited in the disk gas at the locations where the waves damp. Because $\Omega_{\rm p} > \Omega(r)$ in the disk lying exterior to the planet, the waves propagating there carry positive angular momentum relative to the disk material. Hence they exert a positive torque on the disk when they damp. To conserve angular momentum, their excitation must therefore remove angular momentum from the planet. The opposite is true of the waves that are excited at the inner Lindblad resonances, and hence we see that the inner and outer Lindblad resonances exert torques of different signs on the planet. Normally, the outer Lindblad torque dominates because the outer Lindblad resonance lie closer to the planet than the inner ones \citep{Ward1986, Ward1997}. Hence the total Lindblad torque is normally negative, causing the planet to lose angular momentum and spiral into the star. 

We note that as $m \rightarrow \infty$, the resonances appear to get closer and closer to the planet's orbital location (see eqn.~\ref{eqn:Omega_LR}). In a gas disk, however, acoustic waves cannot be launched in the region close to the orbital radius of the planet, where the relative motion of the gas and the planet is subsonic. This is a general feature of fluid flow around an obstacle: waves are not excited in such a flow when the fluid motion relative the the obstacle is subsonic \citep{LandauLifshitz}. The motion of the gas relative to the planet becomes sonic at a distance of $2H/3$ from the planet, and the resonances that lie beyond this distance supply most of the Lindblad torque (in particular those with $m \sim r/H$). The resonances that sit interior to $2H/3$ provide only a small contribution to the torque -  a phenomenon known as the torque cut-off for modes with $m \gg1$\citep{GoldreichTremaine1980,Artymowicz1993}. The superposition of the waves launched at Lindblad resonances give rise to a one-armed spiral density wave \citep{OgilvieLubow2002}, which is often referred to as \emph{the wake}. This feature is shown in Fig.~\ref{fig:spiral-horseshoe}.

It is possible to solve the wave excitation problem using linear perturbation theory, and to calculate analytically the Lindblad torque using the WKB approximation \citep[e.g.][]{GoldreichTremaine1979}. It is also possible to solve the problem numerically. Treating the disk as a 2D structure, with softened gravity included to account for 3D effects, \cite{Paardekooper2010} obtained the following expression for the total Lindblad torque
\begin{equation}
\Gamma_{\rm L} = \gamma (-2.5 - 1.7 \beta + 0.1 \alpha) \Gamma_0
\label{eqn:Gamma_LR}
\end{equation}
where $\alpha$ and $\beta$ are the power-law indices for the disk surface density and temperature, and $\gamma$ is the ratio of the specific heats. $\Gamma_0$ sets the magnitude of the Lindblad torque and is given by
\begin{equation}
\Gamma_0 = \frac{q^2}{h^2} \Sigma_{\rm p} r_{\rm p}^4 \Omega_{\rm p}^2,
\label{eqn:Gamma_0}
\end{equation}
where $q=m_{\rm p}/M_*$ and $h$ is the disc aspect ratio (defined above in the section describing disc models). Symbols with a subscript ``p" are evaluated at the planet's location. We can see from eqns.~(\ref{eqn:Gamma_LR}) and (\ref{eqn:Gamma_0}) that the \emph{net} torque acting on a planet due to the competition between inner and outer Lindblad resonant torques is equal to $\Gamma_0$ multiplied by a factor of order unity. We can ask \emph{where does the form for $\Gamma_0$ come from}? Early investigations of migration torques considered the torque arising from just one side of the disc \citep{GoldreichTremaine1979,LinPapaloizou1979}, and this one-sided torque has the form $\Gamma_{1} \sim \Gamma_0/h$, and so the net torque differs from the one-sided torque by a factor of $h$. The one-sided torque can be derived using a simple picture of fluid elements shearing past the planet and being deflected by the planet's gravity using the impulse approximation \citep{LinPapaloizou1979}. If the minimum distance of approach to the planet by the fluid elements is taken to equal $H$, the disc scale height (justified by our discussion above about the torque cut-off), then an expression for the one-sided torque of the form given by $\Gamma_{1}$ above is obtained. The net torque, $\Gamma_0$, arises from the difference between the inner and outer Lindblad torques, and this difference originates in the fact that outer Lindblad resonances naturally lie closer to the planet than inner Lindblad resonances, an effect that is magnified by the fact that pressure-support in the disc causes the rotation profile to be slightly sub-Keplerian \citep{Ward1997}. This gives rise to an asymmetry between inner and outer Lindblad resonances that scales as $\sim h$, explaining the ratio between $\Gamma_1$ and $\Gamma_0$.

The time scale for migration can be obtained from eqn~(\ref{eqn:dadt}), setting $\Gamma \simeq \Gamma_0$:
\begin{equation}
\tau_{\rm mig} = \frac{r_{\rm p}}{| d r_{\rm p}/dt |} = \frac{1}{2} \frac{h^2}{q^2} \frac{m_{\rm p}}{\Sigma_{\rm p} r_{\rm p}^2} \Omega_{\rm p}^{-1}.
\label{tau_mig}
\end{equation}
An Earth-mass planet located at 1 AU would migrate inwards on a time scale of $\sim 1 .7 \times 10^5$ years, while for a Neptune-mass body  $\tau_{\rm mig} \sim10^4$ years. These time scales are much shorter than the expected disk life times of $10^6$ - $10^7$ years. The question of how to stop or slow down the migration induced by the Lindblad torques has been the focus of much work in the last few years, and is discussed below.
\subsubsection{Corotation torque}
The corotation torque arises through interaction between a planet and material in the disk that corotates with the planet on average. Hence, the disk material involved follows orbits that sit close to those of the planet. The first calculations of the corotation torque were based on solving the linearised disk equations, using methods similar to those used to evaluate the Lindblad torque discussed above \citep[e.g.][]{GoldreichTremaine1979}, and we will refer to estimates of the corotation torque obtained by these methods as the \emph{linear corotation torque}. A key result is that for an isothermal disk, the corotation torque scales as the local gradient of the inverse vortensity, where the vortensity is defined by
\begin{equation}
\omega = \frac{\nabla \times {\bf v}}{\Sigma} = \frac{\Omega}{2 \Sigma} \,\,{\rm (for \,\, a \,\, Keplerian \,\, disk)},
\label{eqn:vortensity}
\end{equation}
and the second equality applies to a Keplerian disk. More recently, it was realised that in a near-adiabatic disk (i.e. one in which cooling is slow), the corotation torque is also dependent on the gradient of the entropy \citep{PaardekooperMellema2006, BaruteauMasset2008, PaardekooperPapaloizou2008}. For disks with power-law surface density and temperature profiles, the power-law index of the entropy is given by $\xi=(\gamma-1)\alpha - \beta$. Estimates of the linear corotation torque show that it is positive (and hence works in the opposite sense to the Lindblad torque, slowing down inward migration) for disk models that have shallow surface density profiles with $\alpha < 3/2$, and for which the radial entropy gradient is negative (i.e. negative $\xi$). In general, however, the linear corotation torque is too weak to significantly slow down migration due to the Lindblad torques. 

We may ask the question \emph{why is it that the vortensity and entropy gradients are fundamental is determining the corotation torque?} The answer is that both of these quantities are conserved on streamlines in ideal disc models, and hence when fluid elements move from one radius to another due to interaction with the planet, the requirement that local pressure equilibrium is maintained, and vortensity and entropy are conserved on streamlines, leads to the development of density perturbations that provide a torque on the planet. More specifically, vortensity is conserved on streamlines in a 2D inviscid barotropic disk (i.e. one where the equation of state is a function of density only), and entropy is conserved on streamlines in an adiabatic disc,  where there is no heat exchange. Conservation of both quantities requires that there are no shocks in the disc. As will be discussed below, some deviation is required from these idealised disc models in order that a corotation torque can be maintained over long time scales against the effects of saturation. Hence, the above arguments also apply in viscous discs that support heat transport under certain limits.

An alternative expression for the corotation torque in an isothermal disk was obtained by \cite{Ward1991} by considering that fluid elements orbiting close to the planet's orbital radius undergo horseshoe orbits relative to the planet (see Fig.~\ref{fig:spiral-horseshoe}). We will refer to estimates of the corotation torque obtained from this type of analysis as the \emph{horseshoe torque}, $\Gamma_{\rm HS}$. Horseshoe orbits do not appear in a linear analysis, and so a torque estimate based on considering the exchange of angular momentum between a planet and fluid elements undergoing horseshoe turns goes beyond linear theory. Nonetheless, the horseshoe torque shows the same scaling with the gradient of the vortensity as the linear corotation torque. Using radiation-hydrodynamic simulations, \cite{PaardekooperMellema2006} showed that the torque experienced by a planet in a near-adiabatic disk with a negative entropy gradient could be strongly positive, demonstrating that the horseshoe torque also depends on the entropy gradient in the disk \citep[see also][]{BaruteauMasset2008,PaardekooperPapaloizou2008}. A simple physical explanation for the origin of the entropy-related horseshoe torque can be given as follows.

\begin{figure}
\includegraphics[scale=.25]{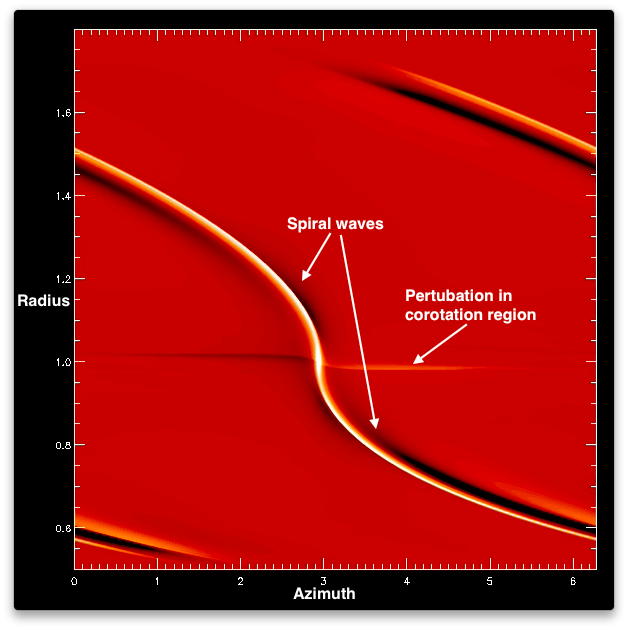}
\includegraphics[scale=.35]{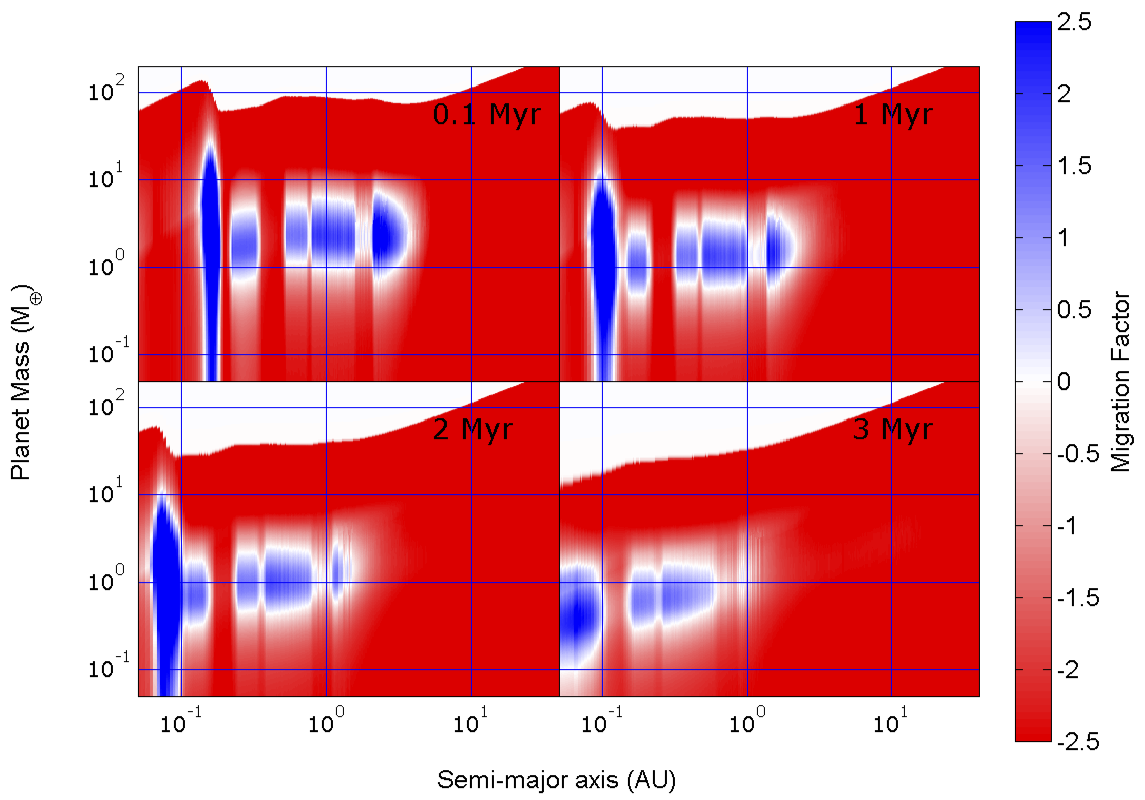}
\caption{Left panel: Close-up view from 2D hydrodynamic simulation of 6 M$_{\oplus}$ planet embedded in a viscous disk, showing spiral waves and density perturbations in horseshoe region arising from the vortensity and entropy gradient. Right panel: Evolution of migration contours as disk evolves. Blue contours show regions of outwards migration. Red regions show regions of inwards migration. White contours show parameters for which the sum of the negative Lindblad torque and the positive corotation torque lead to zero net torque acting on the planet. Similar figures to this are discussed in detail in \cite{ColemanNelson2014} and \cite{ColemanNelson2016}.}
\label{fig:migrationmap}       
\end{figure}

Consider a disk with a negative entropy gradient (i.e. with $\xi$ being negative), such that the temperature decreases strongly with orbital radius. Figure~\ref{fig:spiral-horseshoe} shows that colder gas orbiting just outside the planet's orbital radius will undergo a horseshoe U-turn in front of the planet. This gas conserves its entropy as it moves onto an orbit just interior to the planet, where the ambient gas is warmer. In order to preserve pressure equilibrium the colder fluid elements must increase their density, so a high density region must develop just in front of the planet, as shown in the left panel of Fig.~\ref{fig:migrationmap}. The opposite happens to the warmer fluid elements that make a U-turn just behind the planet, leading to a low density region developing there. This density asymmetry leads to a positive torque. A similar process arises because of the conservation of vortensity during horseshoe U-turns, so the existence of both entropy and vortensity gradients in the disk can drive strong corotation torques. 

Using a 2D disk model with softened gravity, \cite{Paardekooper2010} obtained the following expression for the horseshoe torque \citep[see also][for similar expressions]{MassetCasoli2010}
\begin{equation}
\Gamma_{\rm HS} = \gamma\left[1.1 \left(\frac{3}{2} - \alpha\right)  + 7.9 \frac{\xi}{\gamma} \right] \Gamma_0.
\label{eqn:Gamma_HS}
\end{equation}
The first term represents the vortensity related part, and the second term the entropy related part. Evaluating both eqns.~(\ref{eqn:Gamma_LR}) and (\ref{eqn:Gamma_HS}) shows that the horseshoe drag can be positive and exceed the Lindblad torque in disks with negative entropy gradients and where the surface density profile is flat or increases outwards, and hence migration can be stopped or even reversed by the horseshoe torque \citep[e.g.][]{Masset2006}.

Life is not simple, however, and we must now consider the fact that the horseshoe torque can saturate and essentially switch off under some circumstances. This occurs because the horseshoe orbit period is different for fluid elements undergoing horseshoe orbits at different distances from the planet's orbital radius. Hence, over time the horseshoe region becomes phase mixed and the original gradients in entropy and vortensity that drive the corotation torque are flattened out, switching off the corotation torque. We require thermal evolution of the corotating gas to occur on a time scale that is shorter than the horseshoe libration period (but longer than the time taken for the U-turns to occur) in order for the entropy gradient to be maintained at the optimal level. And we require viscosity to re-establish the surface density and vortensity gradients on the same time scale. The effects of saturation are accounted for in the migration torque prescriptions that are given in \cite{Paardekooper2011}. Figure~\ref{fig:migrationmap} shows the migration behaviour as a function of planet mass and orbital radius in a typical disk model that is heated by viscosity and irradiation from the central star. We see that outwards migration is favoured in the regions close to the star where viscous heating dominates and a negative entropy gradient is established. In the outer irradiation dominated regions, however, migration is dominated by the Lindblad torques and is directed inwards. Planet traps, or zero-migration zones, are also present where Lindblad and corotation torques sum to zero, and such features have been discussed under various contexts in the literature \citep[e.g.][]{Masset2006, LyraPaardekooper2010,HasegawaPudritz2011}. As the disk evolves towards lower mass, the balance of Lindblad and corotation torques changes with the changing disk conditions, and the migration contours in Fig.~\ref{fig:migrationmap} shift both downwards in mass and inwards towards the central star, such that on time scales corresponding to the disk lifetime all planets are expected to drift towards the central star \citep{LyraPaardekooper2010}. The consequences of this and other effects for planet formation are discussed in \cite{ColemanNelson2014, ColemanNelson2016}.

\subsection{High mass planets}
In our discussion so far, we have considered the fact that a planet embedded in a disk can excite spiral density waves at Lindblad resonances, which propagate away from the resonances as acoustic waves. We now consider what happens when these waves deposit their angular momentum in the disk through wave dissipation.

Let us consider the part of the disk that orbits exterior to the planet's orbit. Spiral waves excited at outer Lindblad resonances carry positive angular momentum, and when they damp the disk gas gains angular momentum and moves to a larger orbit. Hence, the deposition of angular momentum in the disk can lead to the redistribution of mass and the formation of a gap in the disk. In a viscous disk, the torque due to viscosity attempts to close the gap, and hence there is a competition between viscous and tidal torques.

A simple estimate for the onset of gap formation can be obtained as follows. The one-sided torque exerted by a planet on a disk can be written as \citep{PapaloizouLin1984}
\begin{equation}
\Gamma_{\rm p} \approx 0.23 q^2 \Sigma_{\rm p} r_{\rm p}^4 \Omega^2_{\rm p} \left(\frac{r_{\rm p}}{\Delta} \right)^3  
\label{eqn:Gamma_p}
\end{equation}
where $\Delta$ is the distance between the planet and the location in the disk where the torque is applied. Note that this one-sided torque is equivalent to $\Gamma_1$ that was introduced during our discussion above concerning the origin of $\Gamma_0$ in eqn.~(\ref{eqn:Gamma_0}). In a Keplerian disk, the viscous torque that opposes the tidal torque is given by
\begin{equation}
\Gamma_{\nu} = 3 \pi \nu \Sigma r_{\rm p}^2 \Omega_{\rm p}.
\label{eqn:Gamma_nu}
\end{equation}
If $q > (H/r_{\rm p}^3)$, then the spiral wake is nonlinear when launched, and will dissipate as a shock close to the planet on a scale of $H$ \citep{Ward1997,KorycanskyPapaloizou1996}. (The planet mass corresponding to $q=(H/r_{\rm p})^3$ is often referred to as the \emph{thermal mass}, $M_{\rm th}$.)
Furthermore, if we adopt the $\alpha$ model for viscosity, $\nu = \alpha_{\rm ss} H^2 \Omega$, then a gap of width $H$ can be maintained against viscous spreading if
\begin{equation}
q \ge 40 \alpha_{\rm ss} \left(\frac{H}{r_{\rm p}}\right)^2.
\label{eqn:q_gap}
\end{equation}
Adopting the disk parameters described above at 1 AU, a planet which exceeds the thermal mass and satisfies eqn.~(\ref{eqn:q_gap}) will have a mass $m_{\rm p} \ge 20 \; {\rm M}_{\oplus}$. Further out in the disk at $a_{\rm p} \sim 5$ AU, where the disk thickness is expected to have a larger value of $h \sim 0.05$, gap opening occurs for $m_{\rm p} \ge 40 \; {\rm M}_{\oplus}$.

The picture of gap formation used to obtain eqn.~(\ref{eqn:q_gap}) is useful to orientate our thinking, but is too simple because gap formation is a continuous process. Even for a massive planet such as Jupiter, some of the torque from the planet is carried into the disk in the form of spiral waves, giving rise to a `pressure torque' that can modify the gap opening criterion \citep{Crida2006}. Furthermore, gaps are observed to start forming in hydrodynamic simulations for masses that are smaller than those predicted by eqn.~(\ref{eqn:q_gap}) and the requirement that the planet mass exceeds the thermal mass. As we will discuss below, spiral waves can damp because of viscosity \citep[e.g.][]{Takeuchi1996} or because they steepen into shocks after travelling a distance of only a few $\times H$ \citep{GoodmanRafikov2001}, and hence can start to influence the density profile close to the planet as they propagate and deposit their angular momentum in the disk, even if they are not launched as shocks. For the purpose of illustration only, consider the 1D diffusion equation for the evolution of the surface density in a viscous, Keplerian disk \citep[][see also the chapter in this volume by Armitage]{Pringle1981}, but modified to include the influence of the torque due to a planet \citep[e.g.][]{Varniere2004}
\begin{equation}
\frac{\partial \Sigma}{\partial t} = \frac{3}{r} \left[r^{1/2} \frac{\partial }{\partial r} \left(\nu \Sigma r^{1/2} \right) -\frac{1}{3 \pi \Omega r}\frac{\partial \Gamma_{\rm p} }{\partial r} \right].
\label{eqn:dsigmadt}
\end{equation}
Note that here $\Gamma_{\rm p}$ should \emph{not} be interpreted as having the exact form given by eqn.~(\ref{eqn:Gamma_p}), but rather should be seen as the torque that is \emph{deposited} in the disk as a function of distance from the planet, through the dissipation of the spiral waves. Hence the precise form of $\Gamma_{\rm p}$ to be used in eqn.~(\ref{eqn:dsigmadt}) will depend on where the spiral waves are excited and dissipate. If we consider a steady solution ($\partial \Sigma/\partial t=0$) and assume $\nu={\rm constant}$, then we can write
\begin{equation}
\frac{\nu \Sigma}{2} = - r \nu \frac{d \Sigma}{dr} + \frac{1}{3 \pi \Omega r} \frac{d\Gamma_{\rm p}}{dr} + \,\, {\rm constant}.
\label{eqn:nu-Sigma}
\end{equation} 
Far from the planet the mass flux through the disk is given by $\nu \Sigma/2 \approx {\rm constant}$, but close to the planet the first two terms on the right hand side combine to determine the structure of the disk and any gap that forms. Hence, given a form for the wave excitation and dissipation, such that $\Gamma_{\rm p}$ is known, and a viscosity law, one can determine the depth and structure of any gap that forms. 

A number of recent studies have been undertaken to examine gap formation \citep{DuffellMacFayden2013,Fung2014,Duffell2015,Kanagawa2015,Kanagawa2017}. In particular, \cite{Duffell2015} adopted the wave damping prescription for weakly nonlinear shocks from \cite{GoodmanRafikov2001} and \cite{Rafikov2002a}, and found decent agreement between simulations and a simple analytic formula for the depths and radial structures of gaps in the shallow gap regime, but the agreement was found to be less good for deep gaps. A study specifically aimed at deep gaps was undertaken recently by \cite{Kanagawa2017}, who obtained an empirical formula for wave excitation and propagation in the presence of deep gaps that allowed a 1D theory to be developed that agrees well with hydrodynamic simulations.

The migration of a planet in a deep gap is called Type II migration. In the idealised situation where gap formation arises because of complete tidal truncation of a viscously-evolving disk, flow across the gap is prevented, and the gap region tends towards a very low density. Tidal torques are then applied by material at the gap edges, which repel the planet towards the centre of the gap, and the planet's migration is controlled by the rate at which the disk viscously accretes onto the star \citep{LinPapaloizou1986}. Early hydrodynamic simulations for Jupiter mass planets migrating in a disk with properties similar to the minimum mass nebula supported the idea that giant planets migrate at approximately the viscous rate \citep[e.g.][]{Nelson2000}. 

In reality, tidal truncation does not prevent the flow of material across the gap, and the gap is not completely empty. Tidal torques may then cause migration at a rate that deviates from the viscous flow speed, and recent simulations performed for a broad range of planet masses and disk conditions indicate that this is indeed the case \citep{Duffell2014, DurmannKley2015}, with migration occurring both faster and slower than the viscous rate, depending on model parameters. For example, in simulations with Jovian mass planets embedded in disks with different masses, \cite{DurmannKley2015} find migration speeds of $\sim 1/2$ the viscous flow speed when the local disk mass $a_{\rm p}^2 \Sigma_{\rm p} \sim 0.1 M_{\rm Jup}$, but migration speeds of $5 \times$ the viscous flow speed when $a_{\rm p}^2 \Sigma_{\rm p} \sim 5 M_{\rm Jup}$, indicating that the gas in the gap can be important for setting the migration rate.

\subsection{Migration of intermediate mass planets}
Up to now, we have only discussed circumstances where the migration rate of a planet is determined by an instantaneous disk torque that depends only on the local properties of the disk, but not on the rate of migration itself. There are situations, however, where a feedback loop can occur, and the migration of the planet affects the torque that drives the future orbital evolution. We will discuss such situations that apply to low mass planets in inviscid and advective disks below, but here we focus our brief discussion on intermediate mass planets in viscous disks, since these are the types of planets for which migration feedback has been shown to be important in these disk models \citep{MassetPapaloizou2003}.

\subsubsection{Runaway migration}
Here we consider the situation where a planet migrates inwards due to Lindblad torques. As the planet migrates, most of the material executing horseshoe orbits is trapped in the corotation region, and gets carried along with the planet. Hence, the planet must exert a negative torque on this material in order to drag it along as it migrates, where the torque depends on the planet's migration rate. Through the principle of action-reaction, the horseshoe material must exert a migration-dependent positive torque on the planet that opposes the original direction of migration. The net drift of the planet relative to the surrounding disk gas, however, allows some material outside the horseshoe region to execute a single horseshoe U-turn relative to the planet, exchanging angular momentum with the planet as it does so. For an inwardly migrating planet, this material originates in the inner disk and is propelled directly into the outer disk, and hence removes angular momentum from the planet. It should be clear that the negative torque arising from the orbit-crossing material depends on the mass flow rate across the orbit, and hence depends on the migration rate. This \emph{flow-through torque} provides a positive feedback on the inwards migration.

If the surface density profile of the disk is unaffected by the planet, then the two opposing torques acting on the planet described above cancel out. If a partial gap forms around the planet, such that the corotation region becomes partially depleted, then the torques no longer cancel, and a migration rate dependent net torque exists. To determine whether or not the feedback on migration leads to a runaway, one needs to consider two quantities. The first is the \emph{coorbital mass deficit}, denoted $\delta M$. This is the mass that would need to be added to the corotation region so that the surface density there is equal to the average surface density of the orbit-crossing flow. The second quantity is the sum of the planet mass and the mass of the circumplanetary disk, ${\tilde M_{\rm p}} = M_{\rm p} + M_{\rm CPD}$. \cite{MassetPapaloizou2003} have shown that when $\delta M < {\tilde M_{\rm p}}$, then the flow-through torque can accelerate the migration, but no runaway occurs. When $\delta M > {\tilde M_{\rm p}}$, however, then migration runs away and can be very rapid indeed.

Runaway migration, or type III migration as it is sometimes called \citep{Peplinski2008}, has been found to operate spontaneously for $\sim$ Saturn-mass planets in viscous disks with masses a few times larger than the minimum mass solar nebula \citep{MassetPapaloizou2003}.  Migration time scales are typically a few tens of orbits. In principle, type III migration can also be directed outwards, but this needs special conditions to occur such as the planet being initiated next to a sharply increasing surface density feature in the disk \citep{Peplinski2008} or a period of outwards migration being imposed on the planet by hand before it is released \citep{MassetPapaloizou2003}. Eventually, all episodes of runaway outwards migration reported so far have been found to stall, such that the planet reverts back to inwards migration.

\section{Migration in inviscid disks}
We now consider the migration of low mass planets in inviscid or very low viscosity disk models. Such models apply when a protoplanetary disk loses angular momentum and accretes onto the central star because of the launching of a magnetised wind from the surface layers, with the midplane regions remaining laminar and experiencing no magnetic torques or radial gas flow. The Lindblad torque discussed previously for viscous disk models is expected to act on a low mass non-gap forming planet similarly in an \emph{inviscid disk}, but the behaviour of the corotation torque is different in this case. We do not make a strong distinction between low and high mass planet migration in the discussion below, because such a distinction does not really apply in an inviscid disk.

\subsection{Dynamical corotation torques}
In the section describing the migration of low mass planets in viscous disks, we discussed the influence of \emph{static} vortensity/entropy-related corotation torques in determining the speed and direction of migration. The influence of the planet's motion through the disk on the migration torque was not discussed in that section, because in a viscously-dominated disk, efficient diffusion of vortensity and entropy is expected to maintain the gradients in those quantities that drive the migration, and the motion of the planet through the disk is not expected to provide a feedback onto the migration torque. The same is not true, however, in an inviscid disk.

As shown by \cite{Paardekooper2014}, in an inviscid disk the migration of the planet can lead to a feedback on the migration torque. The radial motion of the planet through the disk modifies the geometry of the horseshoe orbits, and also introduces streamlines on which fluid elements undergo a single U-turn encounter with the planet while passing from exertior/interior orbits onto interior/exterior orbits, depending on whether planet is migrating inwards or outwards. Considering only the influence of the vortensity gradient in the disk for the time being, the question of whether or not the feedback is positive or negative depends on the direction of migration and on the background vortensity gradient in the disk. The dynamical corotation torque can be written
\begin{equation}
\Gamma_{\rm HS} = 2 \pi \left( 1 - \frac{w_{\rm c}}{w(r_{\rm p})} \right) \Sigma_{\rm p} r_{\rm p}^2 x_{s} \Omega_{\rm p} \frac{d r_{\rm p}}{dt},
\label{eqn:Gamma_dynamical}
\end{equation}
where $w_{\rm c}$ and $w(r_{\rm p})$ are the inverse vortensities measured in the corotation region and in the surrounding disk at the planet's location, respectively. The factor $d r_{\rm p}/dt$ is the planet's migration speed, with a negative sign indicating inwards migration, and hence we see eqn.~(\ref{eqn:Gamma_dynamical}) gives a torque that explicitly depends on the speed and direction of the planet's migration. To illustrate the behaviour of the dynamical corotation torque, let us consider the inwards migration of a planet in a disk with a surface density power law index $\alpha < 3/2$. Here the vortensity in the background disk increases as the planet moves inwards, and hence the inverse vortensity $w(r_{\rm p})$ decreases. Recalling that vortensity is conserved on streamlines in an inviscid disk, the material trapped on librating horseshoe orbits moves with the migrating planet and conserves its vortensity. Hence, the value of $w_{\rm c}$ is set by the location in the disk where the planet begins its migration, and the ratio $w_{\rm c}/w(r_{\rm p})$ starts at unity and increases as the planet migrates inwards. The corotation torque given by eqn.~(\ref{eqn:Gamma_dynamical}) is positive and increases for an inwardly migrating planet, and hence its influence is to gradually cause the inwards migration induced by the Lindblad torques to stall. For disk models with masses of a few times the mass of the minimum mass nebula, and planet masses of a few Earth masses, dynamical torques can induce substantial slowing or stalling of migration long before the planets reach the inner regions of the disk \citep{Paardekooper2014}, and hence these torques can provide a powerful mechanism for slowing the migration of low mass planets.

\subsection{The role of spiral wave dissipation}
In our above discussion about Lindblad torques, it was pointed out that the spiral density waves excited at Lindblad resonances carry an angular momentum flux that is transferred to the disk gas only when the waves dissipate. This wave damping, and deposition of angular momentum in the disk, can be particularly important in inviscid/low viscosity disks as the angular momentum transfer can modify the surface density on either side of the planet, and hence also modify the balance of the inner/outer Lindblad torques that drive migration. In a disk where accretion onto the star is driven by viscosity, the tendency for the damping of spiral waves to modify the disk structure is counterbalanced by viscous diffusion, and any perturbations to the disk density are then expected to be small until the planet becomes massive enough to open a gap - i.e. when the tidal torque exceeds the viscous torque.

The question of how and where the waves damp is clearly important for determining the structure of the disk and the migration behaviour of an embedded planet. Assuming that the spiral waves damp locally at the positions where they are launched, \cite{HouriganWard1984}, \cite{WardHourigan1989} and \cite{Ward1997} examined the combined evolution of the disk surface density and migration of embedded planets. As discussed below, the assumption of local wave damping is probably wrong, and influenced the results obtained in these papers, but nonetheless a number of important insights were obtained in these studies. 

The first was the recognition that an inwardly migrating planet can modify the disk surface density in a way that provides a negative feedback on the migration. To see how this occurs, consider the evolution in a frame of reference that moves with the migrating planet. The gas in the inner disk moves towards the planet in this frame due to the migration of the planet, but near to the planet gas also tends to be repelled from the planet by the damping of the spiral waves that remove angular momentum from the gas. Hence, the gas interior to the planet has a zone of convergence and builds up there. The gas outside the planet appears to move away from it due to the migration, and the addition of angular momentum from the outwards-propagating spiral waves also pushes the material away from the planet. These effects combine to produce a dip in the surface density exterior to the planet. The resulting asymmetry in the surface density distribution either side of the planet acts to increase the torques due to the inner Lindblad resonances and decrease those due to the outer Lindblad resonaces, and hence the net migration torque is reduced, slowing down the migration. 

The second important insight, related to the first, is that there is an upper limit to the planet mass that can continue migrating inwards, known as the \emph{inertial limit} \citep{HouriganWard1984}. For planets above this mass, the feedback effect stalls the migration and a gap is formed in the disk around the position of the planet due to the spiral wave torques. Below this limit, migration continues inwards at a steady rate because the flow of unperturbed gas into the planet's vicinity occurs quickly enough that the surface density asymmetry is maintained at a level that is small enough to prevent migration stalling. Clearly the ability of a planet to form a gap and stall depends on the speed of migration across the gap forming region versus the speed with which a gap is actually formed, and so the mass corresponding to the inertial limit is a function of the local surface density of the disk.

The assumption of local wave damping adopted by \cite{HouriganWard1984}, \cite{WardHourigan1989} and \citep{Ward1997} causes their estimates of the inertial mass to be too small. \cite{Ward1997} gave the critical mass for gap formation and migration stalling to be $M_{\rm crit} \sim 0.2 \Sigma_{\rm p} r_{\rm p}^2 h$, which evaluates to $\sim 0.5$ M$_{\oplus}$ at 1 AU for our disk model. The scaling for the expression for $M_{\rm crit}$ comes from considering the rate at which a planet migrates across the gap region versus the rate at which a gap is formed by the torque applied to the disc by the planet.

Using a local shearing box model, \cite{GoodmanRafikov2001} considered the steepening of spiral waves into weak shocks, and demonstrated that for planets with masses of a few M$_{\oplus}$, linear spiral waves launched by a planet could form shocks within a distance of a few scale heights, $H$, from the planet. Once a shock is formed, the spiral wave amplitude decays as it propagates and its angular momentum is deposited in the disk. The nonlinear shocking length (i.e. distance from the planet where the wave steepens into a shock) was found to be \citep{GoodmanRafikov2001}
\begin{equation}
l_{\rm sh} \sim 0.8 \left(\frac{\gamma+1}{12/5} \frac{m_{\rm p}}{M_{\rm th}}\right)^{-2/5} H.
\label{eqn:l_sh}
\end{equation}
For $m_{\rm p}=M_{\rm th}$ (the thermal mass) and $\gamma=7/5$, this gives $l_{\rm sh} \sim H$, so the wave shocks as it is launched, as expected from our earlier discussion about gap formation in viscous disks. For $m_{\rm p}=M_{\rm th}/10$, we have $l_{\rm sh} \sim 2 H$, so that even for a low mass planet the spiral waves will shock and start to deposit their angular momentum in the disk in the vicinity of the planet. The shock dissipation of the spiral waves was found to give rise to a decay of the angular momentum flux of the waves scaling as 
\begin{equation}
F_{H}(x) \propto | x|^{-5/4} \;\;\;\; (|x| \gg l_{\rm sh})
\label{eqn:F_H}
\end{equation}
where $x = r-r_{\rm p}$ is the distance from the planet. \cite{Rafikov2002a} examined the shock dissipation in global disk models, and the decay of the wave angular momentum flux for this case is shown in his Figs.~1 and 3.

\cite{Rafikov2002b} considered the evolution of migrating planets in inviscid and low viscosity disks under the assumption that the spiral waves dissipate non-locally through shocks. He showed that the same negative feedback mechanism on migration described by \cite{HouriganWard1984} and \cite{WardHourigan1989} also occurrs in this case, and obtained the following estimate for the critical planet mass that could open a gap and stall its migration because of this feedback mechanism
\begin{equation}
M_{\rm cr}= 2.5 \frac{c_{\rm s}^3}{\Omega G} \left(\frac{Q}{h}\right)^{-5/13},
\label{eqn:M_cr2}
\end{equation}
where $Q=\Omega c_{\rm s}/(\pi G \Sigma)$ is the Toomre stability parameter. At 1 AU in the minimum mass nebula $M_{\rm cr} \sim 2$ M$_{\oplus}$, and at 5 AU (assuming $h=0.05$ and $\Sigma =150$ g cm$^{-2}$)  $M_{\rm cr} \sim 8$ M$_{\oplus}$.
Hence, we see that low mass planets can in principle stall their migration and open gaps in inviscid disks, especially if the disk is thin.
\begin{figure}
\includegraphics[scale=1.7]{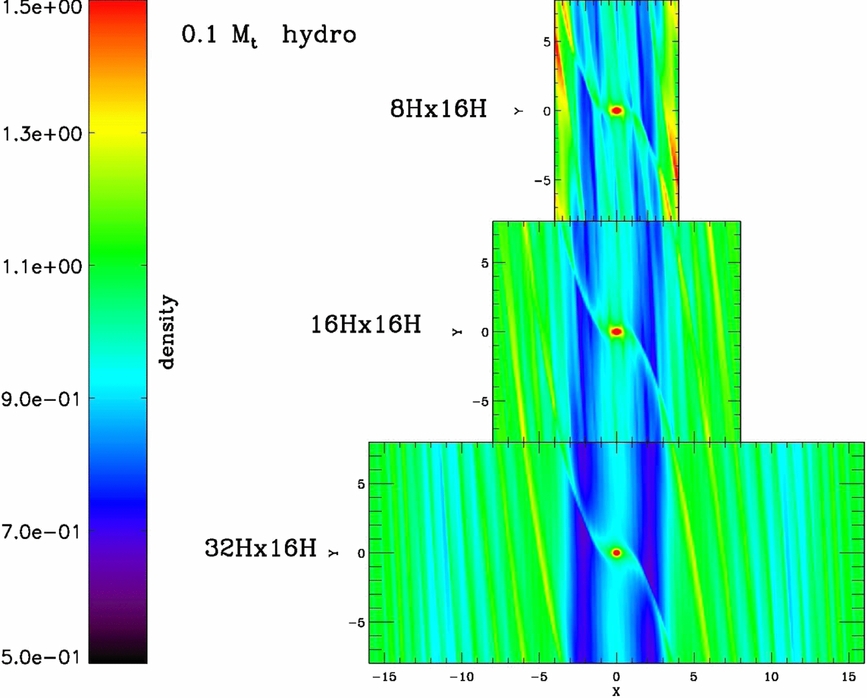}
\includegraphics[scale=1.2]{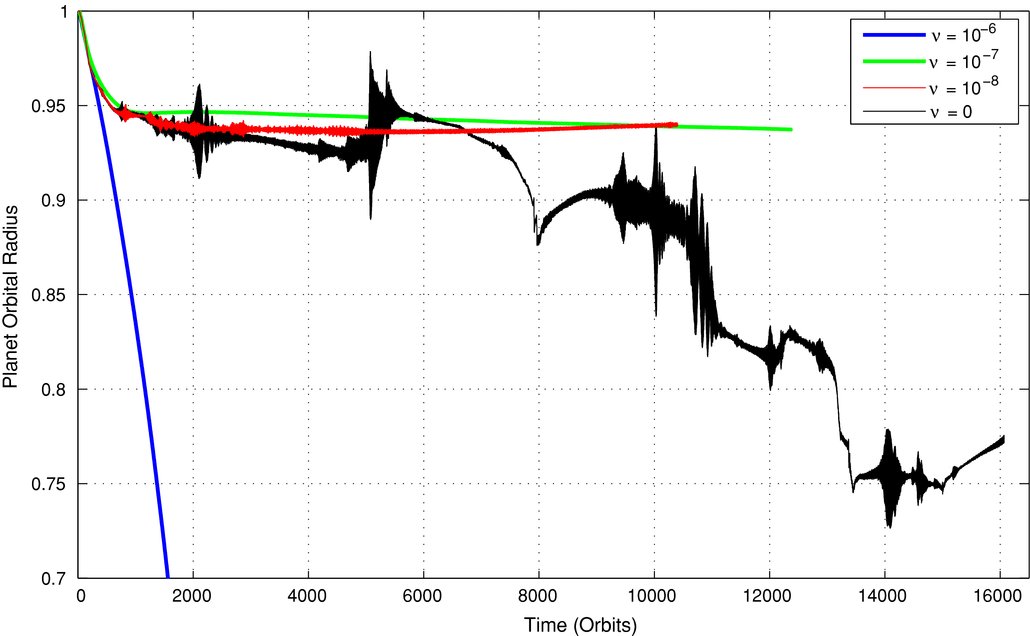}
\caption{Left panel: Results from \cite{Zhu2013} showing gap forming in a disk with a planet mass equal to 10\% of the thermal mass. Right panel: Migration history for 10 M$_{\oplus}$ planets in disks with varying viscosity, showing stalling of migration for low viscosity disks, migration at the expected type I rate in a viscous disk, and erratic migration due to the RWI in an inviscid disk. Taken from \cite{Yu2010}.}
\label{fig:RafikovGap}       
\end{figure}

Detailed hydrodynamic simulations that examined the process of wave steepening and shock dissipation using the local shearing box approximation were undertaken by \cite{Dong2011}. This study showed good agreement between the simulations and the analytic theory in \cite{GoodmanRafikov2001}, and also demonstrated the need for very high resolution for the wave steepening to be captured accurately. Converged results were obtained when 256 cells per scale height were used in the radial domain, but at lower resolutions wave dissipation generally occurred too close to the planet. The process of gap formation for low mass planets was examined in shearing box simulations by \cite{Zhu2013}, and gap formation was demonstrated to occur for $m_{\rm p}=0.1 M_{\rm th}$ (see Fig.~\ref{fig:RafikovGap}). The planets are held on fixed orbits in these simulations, and hence the depths of the gaps formed continue to grow with time as the planet pumps angular momentum into the disk. Global hydrodynamic simulations of live planets in inviscid disks were undertaken by \cite{Li2009}. The disk models chosen had constant vortensity profiles (i.e. $\alpha=3/2$), such that no corotation torque operates. A large suite of simulations undertaken by \cite{Li2009}  showed good agreement with the predictions for $M_{\rm cr}$ given by eqn.~(\ref{eqn:M_cr2}), with the density structure obtained near the planet giving rise to a negative feedback on migration as expected. The long term evolution of planets in disks with varying levels of viscosity was examined by \cite{Yu2010} (see Fig.~\ref{fig:RafikovGap}). Migration at the expected type I rate was obtained for large viscosities $\alpha_{\rm ss} \sim 10^{-3}$. For low viscosities, stalled migration was observed. Interestingly, at zero viscosity, the development of surface density maxima caused by wave damping and gap formation led to the formation of vortices via the Rossby Wave Instability (RWI) \citep{Lovelace1999}, which themselves generate spiral waves and migrate. Interaction between the planet and these migrating vortices led to a very erratic migration behaviour, but one whose net direction appears to be inwards because the vortices themselves migrate inwards. The long term evolution of such a system is clearly something that needs to be explored further. More recently, \cite{FungChiang2017} presented simulations with similar results to \cite{Li2009} and \cite{Yu2010}.

In summary, in inviscid or very low viscosity disks, when the planet mass is below the critical mass for stalled migration and gap formation, we expect that Lindblad torques will drive inwards migration, but dynamical corotation torques will eventually cause migration to stall. If the mass exceeds $M_{\rm cr}$, however, then migration should stall because of nonlinear wave steepening and gap formation.

\section{Migration in advective disks}
We now consider migration in \emph{advective disks}. These are disks in which the flow near the midplane remains laminar, and magnetic stresses there induces radial gas flows. As discussed above, one scenario where such a flow can occur is in a magnetised disk in which the Hall effect generates horizontal magnetic fields that diffuse to the midplane, and wind up to become spiral fields. Such a configuration will induce radial inflow of matter. Other, more complex field geometries in 3D, however, can result in radial \emph{outflow} near the midplane, so for the sake of completeness the discussion below includes the possibility of either inwards or outwards gas flows. As in the case of viscous and inviscid disks, we expect the Lindblad torque to drive inwards migration in an advective disk. The presence of radial flows near the midplane leads to changes in behaviour of the corotation torque. 

\subsection{Low mass planets}
The migration of low mass planets in advective disks has been considered very recently by \cite{McNally2017} and \cite{McNally2018}, who worked in the limit that the disk surface density is unaffected by the planet, and Ohmic resistivity causes magnetic field perturbations induced by the planet to be rapidly damped. \cite{McNally2017} examined the corotation torque acting on a non-migrating planet held on a fixed circular orbit, with the gas flowing radially past the planet with velocity $v_r$. They noted that for a steady accretion flow, the vortensity of material flowing through the disk remains equal to the Keplerian value at each radius (since for slow radial flow the disk maintains near-Keplerian rotation and vorticity profiles), whereas the vortensity of the material trapped on librating horseshoe orbits is continuously modified by the magnetic torque. This evolution of the vortensity in the librating region, however, does not depend on the detailed form of the applied torque, and hence any external torque acting on the gas disc to drive a steady radial accretion flow will have the same effect \citep{McNally2017}. Noting that a planet migrating through a non-viscous, non-accreting disk has a formal similarity to a non-migrating planet in an advective disk, one can deduce that it is the driving of vortensity evolution in the corotation region by the magnetic torque that gives rise to a corotation torque acting on a non-migrating planet. \cite{McNally2017} showed this corotation torque is given by
\begin{equation}
\Gamma_{\rm HS} = 2 \pi \left(1 - \frac{w_{\rm c}(t)}{w(r_{\rm p})}\right) \Sigma_{\rm p} r^2_{\rm p} x_s \Omega_{\rm p} (- v_r),
\label{eqn:Gamma_advective}
\end{equation}
where $w_{\rm c}(t)$ and $w(r_{\rm p})$ are the time-evolving inverse vortensity in the corotation region and the inverse vortensity in the background disk at the planet's location, respectively, and $v_r$ is the radial speed of the gas accretion flow.

We now consider a power-law disk model with $\alpha <3/2$, such that the vortensity increases closer to the star. A negative torque acting on the disk drives inflow and causes the vortensity in the corotation region to increase with time. Hence, the inverse vortensity $w_{\rm c}(t)$ decreases with time, and the corotation torque acting on a non-migrating planet is positive and grows with time according to eqn.~(\ref{eqn:Gamma_advective}). \cite{McNally2017} find that a positive torque acting on the disk leads to an increase of $w_{\rm c}(t)$ and $v_r>0$, and hence the corotation torque is positive in this case too. Ultimately, the positive corotation torque originates from the background vortensity gradient that we have assumed for the disk model, and a disk model with $\alpha>3/2$ would give rise to a negative corotation torque.

\begin{figure}
\includegraphics[scale=.5]{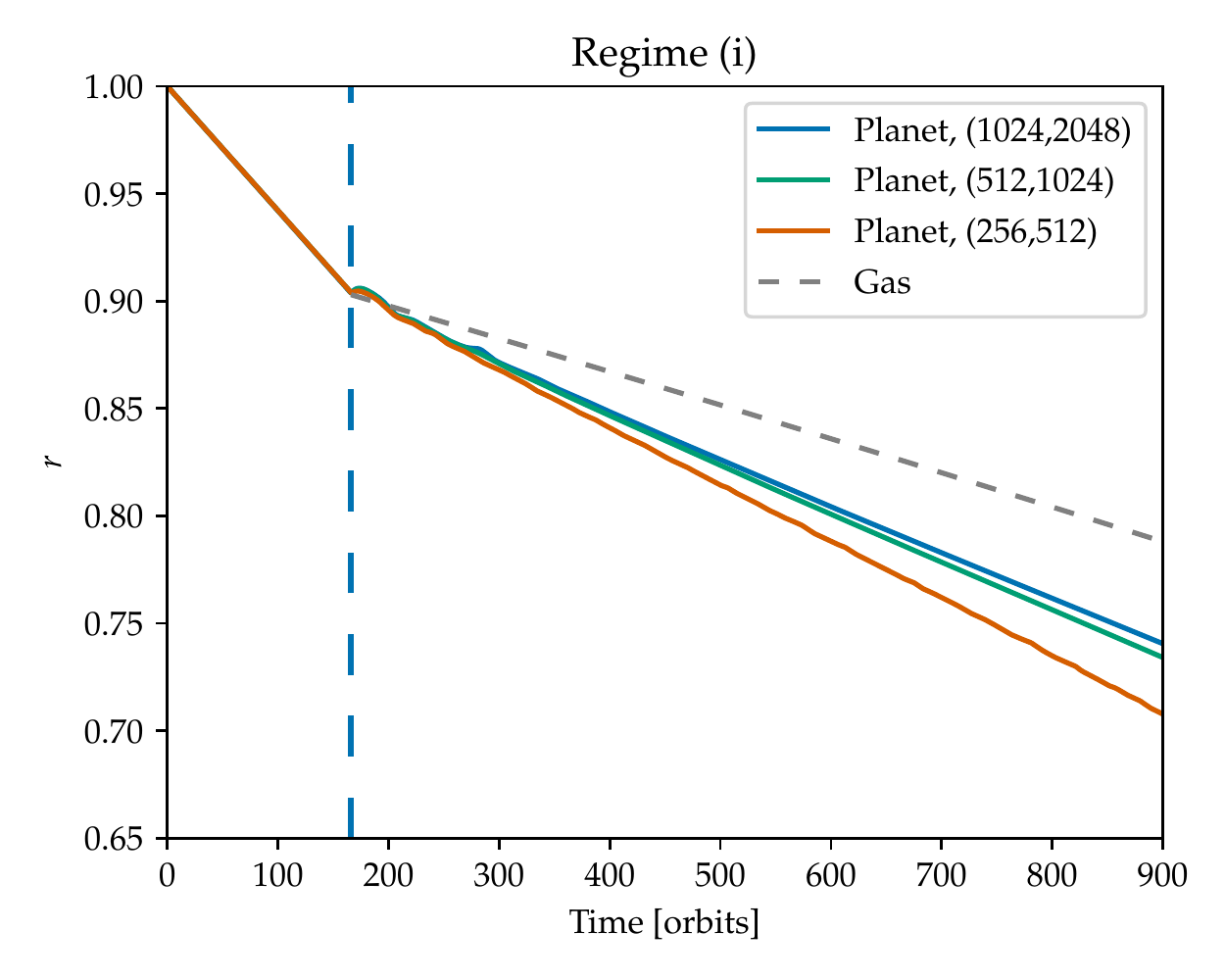}
\includegraphics[scale=.5]{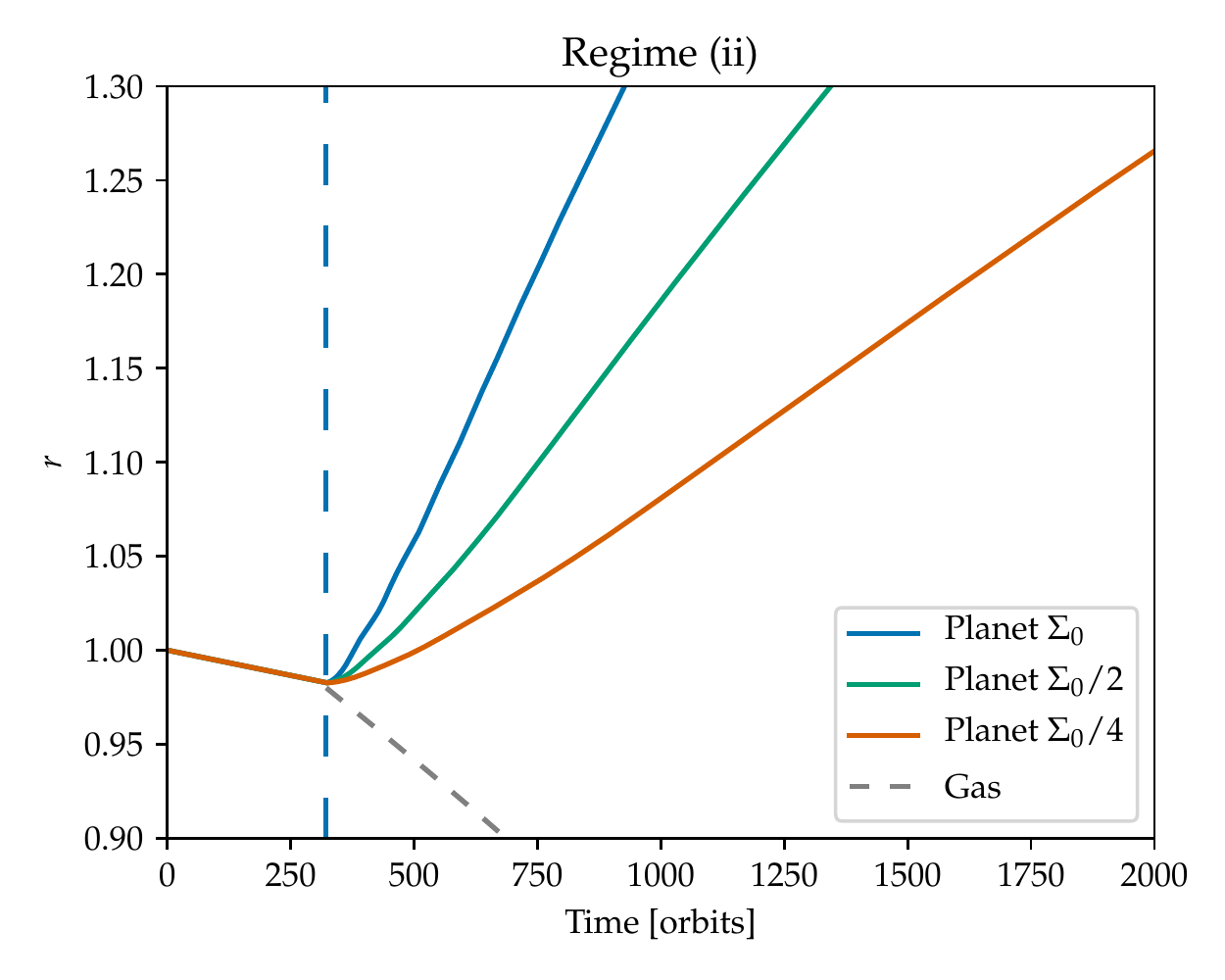}
\includegraphics[scale=.5]{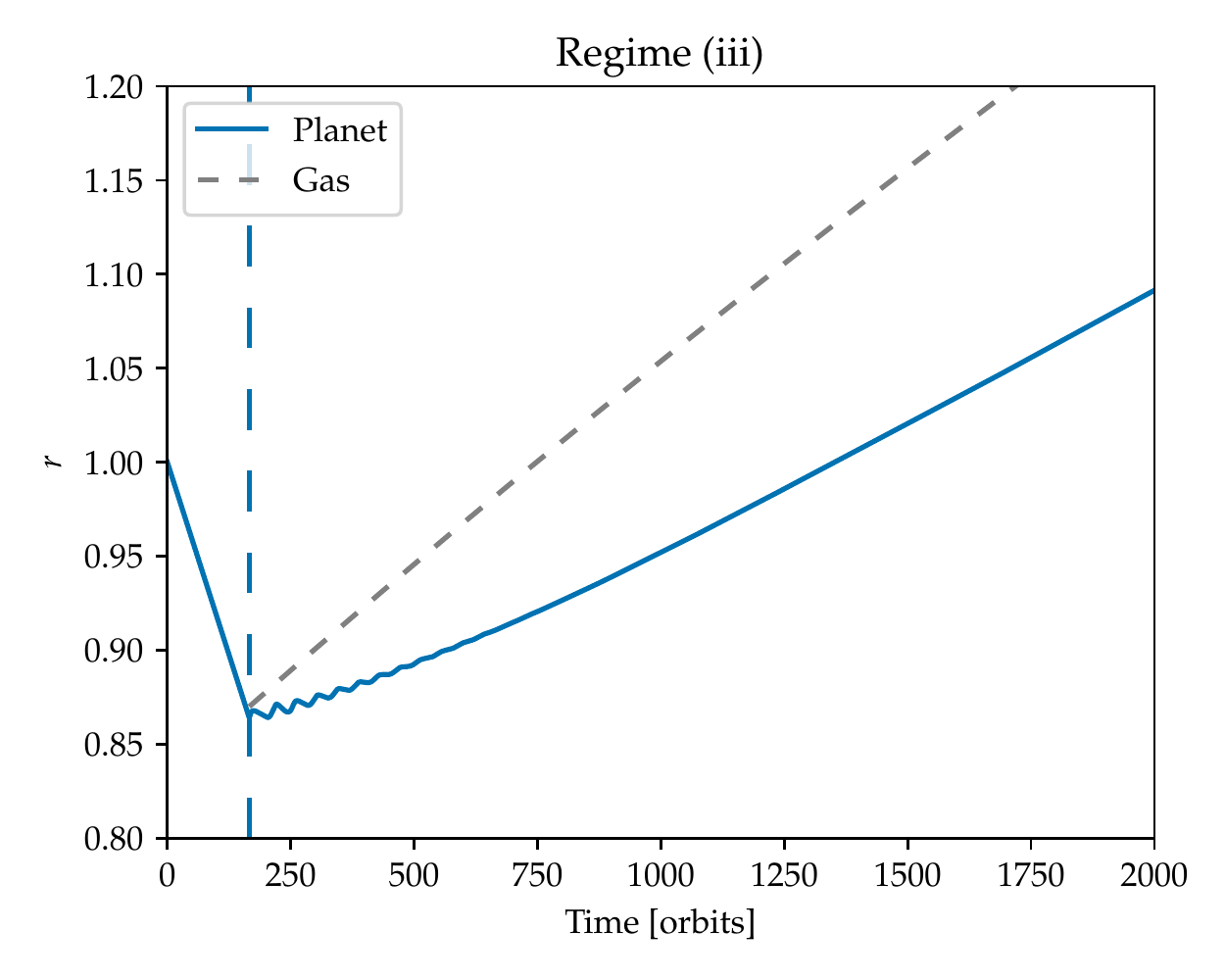}
\includegraphics[scale=.5]{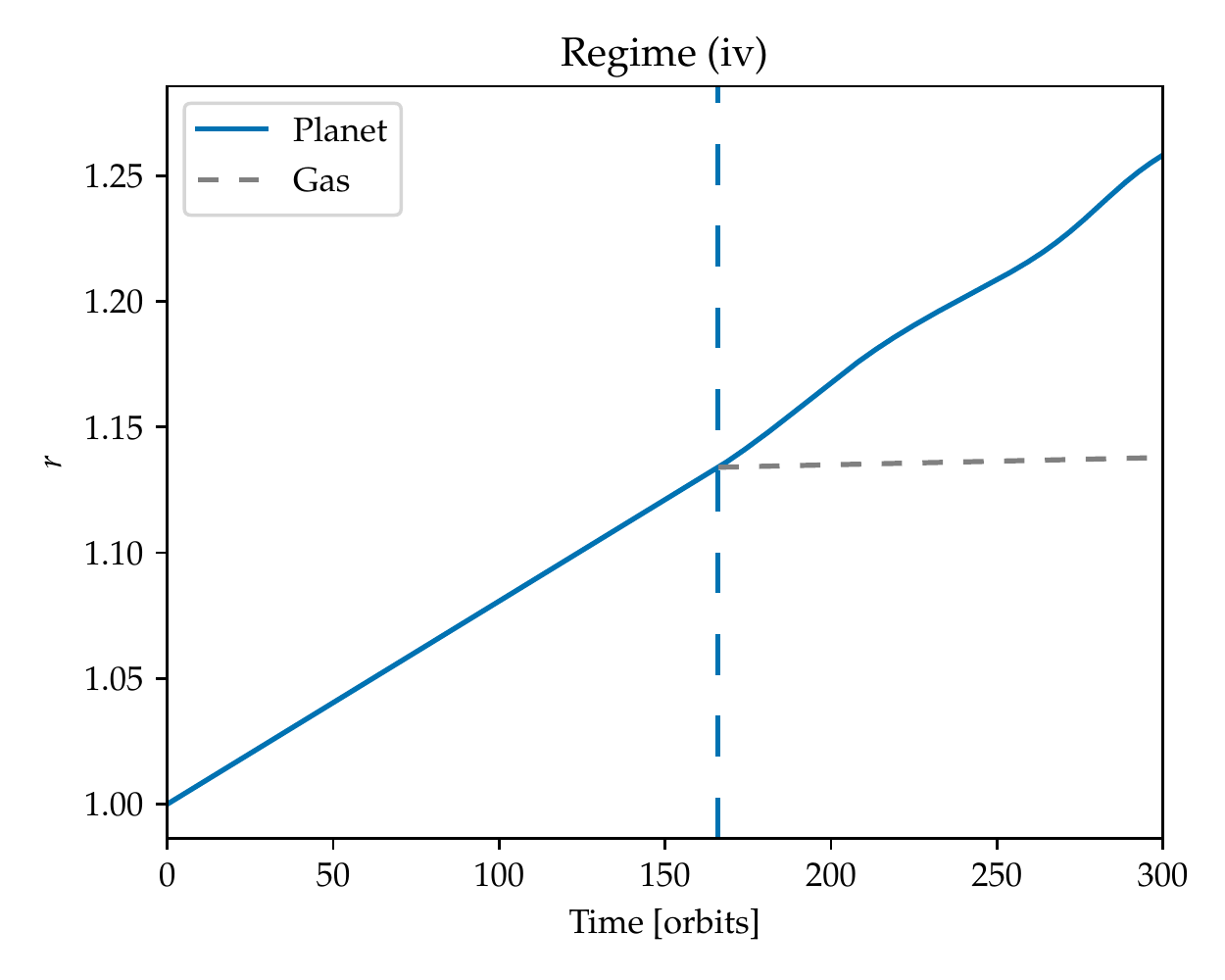}
\caption{Figures taken from \cite{McNally2018} demonstrating the four regimes of migration in advective disks described in the text.}
\label{fig:Regimes}       
\end{figure}

The similarities between eqns.~(\ref{eqn:Gamma_advective}) and (\ref{eqn:Gamma_dynamical}) allowed \cite{McNally2017} to obtain an expression for the corotation torque that applies to a migrating planet in an advective disk:
\begin{equation}
\Gamma_{\rm HS} = 2 \pi \left(1 - \frac{w_{\rm c}(t)}{w(r_{\rm p})}\right) \Sigma_{\rm p} r^2_{\rm p} x_s \Omega_{\rm p} \left[ \frac{dr_{\rm p}}{dt} - v_r \right],
\label{eqn:Gamma_vr+drdt}
\end{equation}
where we recall that $d r_{\rm p}/dt$ is the migration velocity of the planet and $v_r$ is the radial velocity of the gas, with a negative sign indicating inwards motion in both cases.
Analysis of eqn.~(\ref{eqn:Gamma_vr+drdt}) shows that there are four different migration behaviours that are possible, depending on the values that $dr_{\rm p}/dt$ and $v_r$ take. These migration regimes are described below, and their confirmation through simulations was demonstrated by \cite{McNally2018}, and is shown in Fig.~\ref{fig:Regimes}.\\
(i) $v_r <0$, $dr_{\rm p}/dt <0$ and $|dr_{\rm p}/dt| > |v_r|$: Here the disk flow is inwards, the planet is initially migrating inwards driven by the Lindblad torque, and the radial disk flow is slower than the planet migration. Equation~(\ref{eqn:Gamma_vr+drdt}) predicts that the corotation torque will be positive and will attempt to slow the planet down. The corotation torque switches off if $dr_{\rm p}/dt=v_r$, so the planet ends up migrating inwards slightly faster than the gas. Note that in the limit of $v_r=0$ we recover the dynamical corotation torque result discussed above with the planet's migration stalling asymptotically. In the presence of a radial disk flow, this complete stalling no longer occurs and the planet ``surfs along" with the disk flow (as shown in the top left panel of Fig.~\ref{fig:Regimes}).\\
(ii) $v_r <0$, $dr_{\rm p}/dt <0$ and $|dr_{\rm p}/dt| < |v_r|$: Here both the gas and the planet are moving inwards, but the disk now flows faster than the planet. Eqn.~(\ref{eqn:Gamma_vr+drdt}) predicts that the initial inwards planet migration will slow down, stop, turn around and then undergo runaway outwards migration (as demonstrated in the top right panel of Fig.~\ref{fig:Regimes}). \\
(iii) $v_r>0$, $dr_{\rm p}/dt < 0$: Here the gas moves outwards and the planet initially migrates inwards. Equation~(\ref{eqn:Gamma_vr+drdt}) predicts that the planet migration will slow down, stop, reverse, and the planet will migrate outwards at a speed close to that of the outflowing gas. Here the planet ``surfs along" with the outflowing disk material (as shown in the bottom left panel of Fig.~\ref{fig:Regimes}).\\
(iv) $v_r>0$, $dr_{\rm p}/dt>0$, $dr_{\rm p}/dt > v_r$: Here the disk flow is outwards and the planet, for some unspecified reason also initially migrates outwards faster than the disk material. Equation~(\ref{eqn:Gamma_vr+drdt}) predicts that the planet will undergo runaway outwards migration (as shown in the bottom right panel of Fig.~\ref{fig:Regimes}).\\

Inspection of Fig.~\ref{fig:Regimes} shows that for regimes (ii) and (iv), runaway outwards migration is quickly replaced by a long phase of migration at essentially constant speed, depending on the surface density in the disk. \cite{McNally2018} examined the reason for the switch to a constant migration rate, which is not predicted by eqn.~(\ref{eqn:Gamma_vr+drdt}), and showed that at high migration speeds the vortensity perturbation in the corotation region is subjected to ram pressure stripping, breaking the link between increasing migration torque with increasing migration speed required for the runaway to continue. \cite{McNally2018} also considered the reasons for the difference between the migration behaviour of a planet in a viscous disk and that in advective disk, where the disk accretion flow occurs at the same rate in both cases. In an advective disk, the torque that drives accretion also drives evolution of the vortensity in the corotation region, but because there is no mixing between this region and the rest of the disk, a growing contrast between the vortensity in this region and the rest of the disk is maintained. In a viscous disk, however, the viscous diffusion causes the vortensity in the corotation region to mix with the surrounding disk such that the vortensity contrast is smoothed out and becomes essentially time independent, giving rise to the static corotation torque discussed early in this article. 

\subsection{High and intermediate mass planets and gap formation}
To date there have been no studies of gap formation in advective disks, but it should be clear that a significant radial flow through the disk will reduce the tendency for planets to form gaps, particularly when the time for gas to flow across the planet's orbit is short compared to the gap formation time scale. Hence, we would expect the estimate of $M_{\rm cr}$ in eqn.~(\ref{eqn:M_cr2}) to be substantially altered in an advective disk.

\section{Conclusions}
In this review of disk-planet interactions, we have focussed on the role of the disk model in determining the behaviour of a single planet orbiting in a disk. Our discussion has shown that whether a planet is orbiting in a viscous disk, an inviscid disk or a so-called advective disk (where large scale magnetic fields drive a laminar accretion flow through the full column density of the disc) has a profound effect on how it will migrate, and how the disk itself will be shaped by the interaction. Behaviour ranging from rapid migration into the star, to rapid runaway outwards migration over large distances, is possible. In a viscous disc, the direction and speed of migration is determined by the local balance of corotation and Lindblad torques, and one can normally define the instantaneous net migration torque through knowledge of local disc conditions as the torque does not depend on the past history of the planet or on its current migration speed. In an advective disc, however, the situation can be more complicated, and the migration behaviour at any one location and time in the disc can depend on the integrated history of the planet and the torques that have been applied to the disc material to drive the advective accretion flow. Fundamentally, this difference between viscous and advective discs arises because viscosity causes diffusion in the disc, smoothing out the disc properties in the vicinity of the planet, whereas any local contrasts that build up in disc properties between material that is trapped on librating streamlines near the planet and the surrounding disc are maintained in advective discs.  A clear challenge over the coming years will be to develop a consensus view of how the typical protoplanetary disk evolves, and how the variation of disk properties about the mean leads to diversity in the properties of the planetary systems that form and migrate in these disks.

This article has been necessarily limited in its scope, and a number of potentially important issues have been neglected from the discussion. These include some important thermodynamic effects that can influence migration such as the `cold finger effect' \citep{Lega2014}, the `heating torque' induced by irradiation of the disk by a hot planet \citep{Benitez-Llambay2015}, and the role of the entropy gradient in the disk in influencing dynamical corotation torques \citep{Pierens2015}.

Finally, an issue that needs to be considered when comparing migration theory with observations of exoplanets is how the long term evolution of the disk affects planet migration. The emerging picture of disk evolution occurring via layered accretion etc. applies to disks in which ionisation sources are unable to penetrate deeply into the disk. As the disk ages, however, the penetration depth will increase and the nature of the accretion flow is also likely to change. Migration time scales for planets orbiting interior to 1 AU are short, and the planetary configurations that we observe now are likely to have been influenced by disk-planet interactions during this late stage. Hence, in future work it will be important to understand the role of disk-planet interactions throughout the whole of the disk life time.

\begin{acknowledgement}
I am particularly indebted to my colleagues Colin McNally and Sijme-Jan Paardekooper for insightful comments and discussions about disk-planet interactions during the writing of this article.
\end{acknowledgement}

\bibliographystyle{spbasicHBexo}  
\bibliography{Nelson} 

\end{document}